\pdfoutput=1
\documentclass[10pt,pra,aps,superscriptaddress,nofootinbib,twocolumn,longbibliography]{revtex4-1}
\usepackage{amsmath,bbm}
\usepackage{amsthm}
\usepackage{amsmath}
\usepackage{latexsym}
\usepackage{amssymb}
\usepackage{float}
\usepackage{quantikz}
\usepackage{mathrsfs}
\usepackage{tikz}
\usepackage{graphicx}           
\usepackage{color}
\usepackage{xcolor}
\usepackage{mathpazo}
\usepackage{comment}
\usepackage{enumitem}
\usepackage{multirow}
\usepackage{setspace}
\usepackage{subcaption}
\usepackage[colorlinks=true,linkcolor=blue,citecolor=magenta,urlcolor=blue]{hyperref}
\usepackage{svg}

\newcommand{\be}{\begin{equation}}
\newcommand{\ee}{\end{equation}}
\newcommand{\bea}{\begin{eqnarray}}
\newcommand{\eea}{\end{eqnarray}}
\def\squareforqed{\hbox{\rlap{$\sqcap$}$\sqcup$}}
\def\qed{\ifmmode\squareforqed\else{\unskip\nobreak\hfil
\penalty50\hskip1em\null\nobreak\hfil\squareforqed
\parfillskip=0pt\finalhyphendemerits=0\endgraf}\fi}
\def\endenv{\ifmmode\;\else{\unskip\nobreak\hfil
\penalty50\hskip1em\null\nobreak\hfil\;
\parfillskip=0pt\finalhyphendemerits=0\endgraf}\fi}

\makeatletter
\newtheorem*{rep@theorem}{\rep@title}
\newcommand{\newreptheorem}[2]{%
\newenvironment{rep#1}[1]{%
 \def\rep@title{#2 \ref{##1}}%
 \begin{rep@theorem}}%
 {\end{rep@theorem}}}
\makeatother

\newcommand{\cH}{\mathcal{H}}
\newcommand{\Tr}{\operatorname{Tr}}
\newcommand{\cI}{\mathcal{I}}
\newcommand{\cK}{\mathcal{K}}
\newcommand{\cL}{\mathcal{L}}
\newcommand{\cS}{\mathcal{S}}
\newcommand{\cW}{\mathcal{W}}

\newcommand{\Id}{\mathbbm{1}}                 % identity operator
\newcommand{\idch}{\mathrm{id}}               % identity channel
\newcommand{\tr}{\mathrm{Tr}}                 % trace
\newcommand{\Meas}{\mathscr{M}}               % measurements (POVMs)
\newcommand{\Inst}{\mathscr{I}}               % instruments
\newcommand{\Chan}{\mathscr{C}}               % channels
\newcommand{\Bcast}{\mathscr{C}^{\mathrm{br}}} % broadcasting channels

\newcommand{\vb}[1]{\mathbf{#1}}              % real 3-vectors
\newtheorem{theorem}{Theorem}
\newtheorem{proposition}{Proposition}
\newtheorem{observation}{Observation}
\newtheorem{lemma}{Lemma}

\newtheorem{corollary}{Corollary}
\newtheorem{definition}{Definition}
\newtheorem{remark}{Remark}

\begin{document}

\title{Sequential vs parallel Strategies for Sharing Quantum Advantage in Communication Tasks}

\author{Satyaki Manna}
\email{mannasatyaki@gmail.com}
\author{Arindam Mitra}
\email{am56@iitbbs.ac.in; arindammitra143@gmail.com}
\affiliation{Department of Physics, School of Basic Sciences, Indian Institute of Technology Bhubaneswar, Odisha 752050, India.}
\author{Sumit Mukherjee}
\email{mukherjeesumit93@gmal.com}
\affiliation{S. N. Bose National Center for Basic Sciences, Block JD, Sector III, Salt Lake, Kolkata 700106, India.}

\begin{abstract}
Information can be extracted from a quantum system in a sustainable manner, allowing the system to be reused for further information extraction through subsequent interactions. We investigate the possibility of consensus between two experimentally feasible methods, namely, parallel and sequential sharing, that enable multiple observers to extract information encoded in a quantum system. We show that any statistics generated by a parallel sharing strategy can be reproduced by a sequential strategy when only two observers are involved. However, when more than two observers participate in sharing quantum correlations, certain statistics achievable through parallel strategies cannot be reproduced by sequential strategies. Moreover, we demonstrate that sharing quantum advantage through a parallel strategy is impossible for a class of communication tasks involving two decoding measurements, thereby establishing an arbitrarily large gap between the sustainability of quantum advantage under sequential and parallel strategies. This naturally raises the question of whether this impossibility extends to communication tasks involving more than two decoding measurements. We show that, while parallel sharing of quantum advantage is impossible in a 3-bit random access code (RAC), it becomes possible in a 3-bit parity-oblivious RAC, where at most two observers can simultaneously obtain a quantum advantage by performing measurements on two imperfect clones of the initially prepared quantum system. Our findings have potential applications in the development of quantum communication networks and distributed quantum information processing.

\end{abstract}

\maketitle

\section{Introduction}
One of the most distinctive features of quantum interactions is that, unlike their classical counterparts, they are fundamentally invasive in nature \cite{Fuchs96,Fuchs98,busch13}. While this property of quantum theory plays a crucial role in enabling advantages in a plethora of information-processing tasks, particularly quantum cryptography and secure communication \cite{Bennett1984,Fuchs1997,Bennett1992}, it also causes a nuisance in performing certain tasks where information must be extracted repeatedly from the same quantum system. This naturally motivates the search for sustainable ways of implementing quantum interactions, through which quantum advantage can be dispensed among multiple parties.

The conventional method of sharing quantum correlations among multiple observers involves sequential weak measurements, each extracting partial information while preserving sufficient information in the post-measurement state for subsequent observers \cite{silva15,malSharing16,sasmalShar18,Kumari_share19,sasmalPRL24,Tavakoli_sharing_prl,colbeckShareprl,buzekShar11}. Over the past decade, this approach of sharing quantum correlations has found applications in a quite a few number of information-processing tasks \cite{pangPRA13,HilleryDiscrPRL13,PrabuddhaDegeneracy23,AbhyoudaiPRA23,SumitPRA21,Mohan2019,Roy_2023,mitra2026certification,pandit26}. However, comparatively little attention has been devoted to the sustainable sharing of quantum information through alternative methods that do not require sequential weak measurements on the same system. For instance, quantum broadcasting \cite{fuchsBroad96,heinosaari_broad_16,FAN2014241,sarubi,sarubi2026} offers a promising alternative for sharing quantum correlations among multiple observers without requiring sequential measurements on the same system.

Quantum broadcasting, however, is subject to fundamental limitations. Unlike classical information, which can be replicated arbitrarily many times without any loss of fidelity, quantum information cannot, in general, be distributed among multiple systems while preserving the original state in each subsystem, as established by the no-broadcasting theorem \cite{fuchsBroad96}. Despite this limitation, quantum broadcasting holds considerable potential for applications in quantum information processing, owing to its close connections with several fundamental aspects of quantum information theory, including quantum cloning \cite{heinosaari_broad_16,Mitra_2021,arindam_layers21}, measurement incompatibility \cite{Heinosaari_2017,broughamPRA06}, and other nonclassical correlations \cite{HorodechkiPRL08,Uola24,saha15,Bancal09,Chaves17}.

The implementation of quantum information processing tasks in distributed scenarios calls for a systematic and comparative investigation of the available methods for sharing quantum correlations among multiple observers. This motivates us to explore the relative capabilities and fundamental limitations of different approaches to distributing quantum information and nonclassical advantages.

In this work, we introduce \emph{parallel sharing} as an alternative to \emph{sequential sharing} by employing quantum broadcasting to distribute quantum information across multiple systems, thereby enabling multiple receivers to share quantum advantage in a prepare-and-measure scenario. We first show that, when quantum advantage is shared between two receivers, the sequential strategy strictly outperforms its parallel counterpart. In particular, any statistics achievable through a parallel strategy can also be reproduced by a sequential strategy. However, this equivalence no longer holds when more than two receivers are involved. Specifically, we demonstrate that, when the sender aims to share nonclassicality among three receivers, the parallel strategy can generate statistics that cannot be reproduced by its sequential counterpart.

We further investigate the sharing of quantum correlations through parallel strategies in a class of communication tasks characterized by a linear figure of merit. Specifically, we prove that parallel sharing of quantum advantage is impossible in any such communication task where each receiver has only two measurement settings. As a consequence, we identify a class of Random Access Codes (RACs) for which quantum advantage cannot be shared through the parallel strategy. Building on known results \cite{colbeckShareprl,Tavakoli_sharing_prl,sasmalPRL24,sasmal2026}, we further establish the existence of an arbitrarily large gap between parallel and sequential strategies in their ability to share nonclassical correlations in certain RACs. Finally, we demonstrate that, in parity-oblivious random access codes (PORACs), nonclassicality can be shared between two receivers using a universal quantum cloning machine. 

The paper is organized as follows. In the next section (Sec.~\ref{sec:prelim}), we introduce the preliminary concepts and mathematical tools required to formulate the relevant scenarios and establish our results. In Sec.~\ref{sec:compare} we compare parallel and sequential strategies in general. In Sec.~\ref{Sec:two_setting} we investigate the possibility of parallel sharing in a class of communication tasks, while in Sec.\ref{Sec:PM_RAC} we discuss its implication to RACs. Finally, in Sec.~\ref{sec:concl}, we conclude by summarizing our main findings and discussing potential directions for future research.
\section{Preliminaries}\label{sec:prelim}
Throughout this work, we restrict our attention to finite-dimensional Hilbert spaces. For a given Hilbert space $\cH$, we denote by $\cL(\cH)$ the set of linear operators acting on $\cH$, and by $\cS(\cH)\subset\cL(\cH)$ the set of density operators (quantum states). A measurement $M$ on $\cH$ with outcome set $\Omega_M$ is described by a positive operator-valued measure (POVM), defined as a collection $M=\{M(a)\}_{a\in\Omega_M}$ of positive operators with $\sum_{a\in\Omega_M}M(a)=\Id_\cH$. We denote the set of all measurements on $\cH$ by $\Meas(\cH)$. The set of quantum channels (completely positive, trace-preserving maps) from $\cL(\cH)$ to $\cL(\cK)$ is $\Chan(\cH,\cK)$, and $\idch_X$ is the identity channel on system $X$. Further, for a linear map $\Phi: \mathcal{H} \rightarrow \mathcal{K}$, its adjoint map $\Phi^\dagger:\mathcal{K} \rightarrow \mathcal{H}$ is defined by $\tr[\Phi(X)\,Y]=\tr[X\,\Phi^\dagger(Y)]$, with $X \in \mathcal{L}(\mathcal{H})$ and $Y \in \mathcal{L}(\mathcal{K})$.

\emph{Broadcasting of quantum information:-}
In quantum information theory, broadcasting refers to the process of distributing the information encoded in a quantum system among multiple subsystems while preserving the relevant information as much as possible in their local states. As already stated, quantum broadcasting process is constrained by the no-broadcasting theorem \cite{fuchsBroad96}, which states that a set of quantum states can be perfectly broadcast if and only if the states commute. Consequently, unlike classical information, arbitrary quantum information cannot be perfectly distributed among multiple sites by replicating the original state carrying the information into each sites. However, approximate broadcasting is possible, allowing quantum information to be distributed among multiple observers through imperfect copies of the initially prepared system \cite{Buzek1996}. Such a process is mathematically described by a quantum channel, commonly referred to as a broadcasting channel \cite{heinosaari_broad_16,arindam_layers21}, as defined below.
\begin{definition}[Broadcasting channels]
Let $\cH_1=\cdots=\cH_n=\cH$. A channel $\Lambda\in\Chan\big(\cH,\bigotimes_{k=1}^{n}\cH_k\big)$ is called a $1\rightarrow n$ \emph{broadcasting channel}. The set of all such channels is denoted $\Bcast_{1\rightarrow n}(\cH)$.
\end{definition}
We will repeatedly use this broadcasting channel in the subsequent sections to establish our main results.
\subsection{Equivalence of measurement models and quantum instruments}
In scenarios where both the measurement outcomes and the resulting post-measurement states are relevant, quantum measurements are most naturally described within the framework of quantum instruments \cite{Heinosaari_book_2012}. To see this let us consider the general scenario where to perform a measurement $M\in\Meas(\cH)$, the system is coupled to a probe, and the probe is then measured to read off the outcome. Let the system and the probe be initially prepared in the states $\rho\in\cS(\cH)$ and $\omega\in\cS(\cH_E)$, respectively. Let $\mathbf{H}$ be the total Hamiltonian of system and probe, including their interaction, and let $M'\in\Meas(\cH_E)$ be the measurement on the probe, with $\Omega_{M'}=\Omega_M$. We define $U=e^{-i\mathbf{H}t_f}$, where $t_f$ is the time at which the interaction is switched off and the probe is measured. Then, the probability of obtaining outcome $a$ on measurement $M$ is,
\begin{align}
    p(a\,|\,M,\rho)
    &=\tr\big[\rho\, M(a)\big]\nonumber\\
    &=\tr\big[U(\rho\otimes\omega)U^{\dagger}\,(\Id_{\cH}\otimes M'(a))\big],
    \label{Eq:meas_model}
\end{align}
for all $a\in\Omega_M$. The quadruple $(\cH_E,\omega,U,M')$ is called a \emph{measurement model} of $M$.

A \emph{quantum instrument} with input Hilbert space $\cH$ and output Hilbert space $\cK$ is a collection $\cI=\{\Phi_a\}_{a\in\Omega_{\cI}}$ of completely positive maps $\Phi_a:\cL(\cH)\rightarrow\cL(\cK)$ such that $\sum_{a\in\Omega_{\cI}}\Phi_a\in\Chan(\cH,\cK)$, where $\Omega_{\cI}$ is the outcome set of $\cI$. We denote the set of all such instruments by $\Inst(\cH,\cK)$. An instrument $\cI$ induces the measurement $N\in\Meas(\cH)$ given by
\begin{equation}
    N(a)=\Phi_a^{\dagger}(\Id_{\cK}),\qquad a\in\Omega_{\cI},
\end{equation}
so that $\tr[\Phi_a(\rho)]=\tr[\rho\, N(a)]$. Now, given a measurement model $(\cH_E,\omega,U,M')$ of $M$, define for all $a\in\Omega_M$
\begin{align}
    \overline{\Phi}_a(\rho):=\tr_{E}\Big[&\big(\Id_{\cH}\otimes\sqrt{M'(a)}\big)\,U(\rho\otimes\omega)U^{\dagger}\nonumber\\
    &\times\big(\Id_{\cH}\otimes\sqrt{M'(a)}\big)\Big].
    \label{Eq:model_to_inst}
\end{align}
Each $\overline{\Phi}_a$ is completely positive, being a composition of completely positive maps, and $\sum_a\overline{\Phi}_a(\rho)=\tr_E[U(\rho\otimes\omega)U^\dagger]$ defines a channel. Hence $\overline{\cI}:=\{\overline{\Phi}_a\}_{a\in\Omega_M}\in\Inst(\cH,\cH)$, and by Eq.~\eqref{Eq:meas_model}, $\tr[\overline{\Phi}_a(\rho)]=\tr[\rho\, M(a)]$, i.e., $\overline{\cI}$ implements $M$. The post-measurement state for outcome $a$ is
\begin{equation}
    \rho_{a|M}=\frac{\overline{\Phi}_a(\rho)}{\tr[\overline{\Phi}_a(\rho)]}.
\end{equation}
Thus every measurement model defines a quantum instrument. Conversely, by the Stinespring dilation theorem, every instrument admits a measurement model, so the two notions are equivalent. The argument is stated for $\Inst(\cH,\cH)$, but the equivalence holds in general for $\Inst(\cH,\cK)$ as well.
\subsection{Prepare-and-measure scenario}\label{pm_scenario}
 We consider a prepare-and-measure scenario with two spatially separated observers, Alice the sender and Bob the receiver. Alice and Bob receive inputs $x\in[X]$ and $y\in[Y]$, respectively, with $[K]:=\{1,\dots,K\}$ for any positive integer $K$. Upon receiving $x$, Alice prepares a $d$-dimensional classical or quantum system, obtains a local outcome $a\in[A]$ in the process, and sends the system to Bob. After receiving both his input $y$ and Alice's message, Bob produces an output $b\in[B]$. The performance of the task is characterized by the set of conditional probability distributions $\{p(a,b|x,y)\}$.

In classical communication, Alice generates a pair $(a,m)$ from her input $x$ according to the probability distribution $p_e(a,m|x)$ and sends the message $m\in[d]$. Bob then decodes the received message $m$ based on his measurement choice $y$, producing an outcome $b$ according to the conditional probability distribution $p_d(b|m,y)$. Then the resulting statistics can be written as,
\bea\label{c_stat}
p(a,b|x,y)=\sum_{m=1}^{d} p_e(a,m|x)\,p_d(b|m,y).
\eea
The encoding and decoding distributions are non-negative and normalized, i.e.\ $\sum_{a,m}p_e(a,m|x)=1$ for every $x$, and $\sum_{b}p_d(b|m,y)=1$ for every $m,y$.

In quantum communication, Alice and Bob share a bipartite state $\rho_{AB}\in\cS(\cH_A\otimes\cH_B)$ with $\dim(\cH_B)=d$. Alice can in general prepares Bob's local state by measuring her subsystem. For her input $x$, she performs a measurement $N_x:=\{N_x(a)\}_{a=1}^{A}\in\Meas(\cH_A)$, obtains the outcome $a$, which effectively produces the system $B$ of Bob. Subsequently, depending on his input $y$, Bob performs a measurement $M_y:=\{M_y(b)\}_{b=1}^{B}\in\Meas(\cH_B)$ on the received system. Here $N_x(a),M_y(b)\ge0$, $\sum_aN_x(a)=\Id_A$ and $\sum_bM_y(b)=\Id_B$. According to the Born rule, the joint probability distribution of the outcomes obtained by the respective observers is given by,
\begin{align}\label{q_stat}
p(a,b|x,y)&=\tr\left[\rho_{AB}\,\big(N_x(a)\otimes M_y(b)\big)\right] =\tr\left[\rho_{a|x}\,M_y(b)\right],   
\end{align}
where
\bea
\rho_{a|x}:=\tr_A\left[\big(N_x(a)\otimes\Id_B\big)\rho_{AB}\right]
\eea
is the (unnormalized) state prepared on Bob's side when Alice obtains the outcome $a$ for input $x$. The normalized state is $\rho_{a|x}/p(a|x)$ with $p(a|x)=\tr[\rho_{a|x}]$. We consider the ensemble to be no-signaling, i.e., it satisfies $\sum_a\rho_{a|x}=\rho_B:=\tr_A[\rho_{AB}]$, for every $x$.
\subsection{Sequential and parallel strategy}
\begin{figure}%
		\centering
		\subfloat[\centering ]{{\includegraphics[width=9cm]{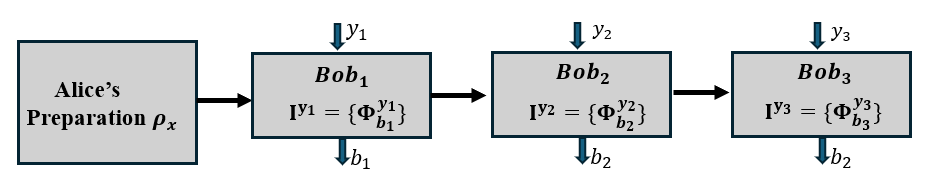} }}%
		\qquad
		\subfloat[\centering ]{{\includegraphics[width=9cm]{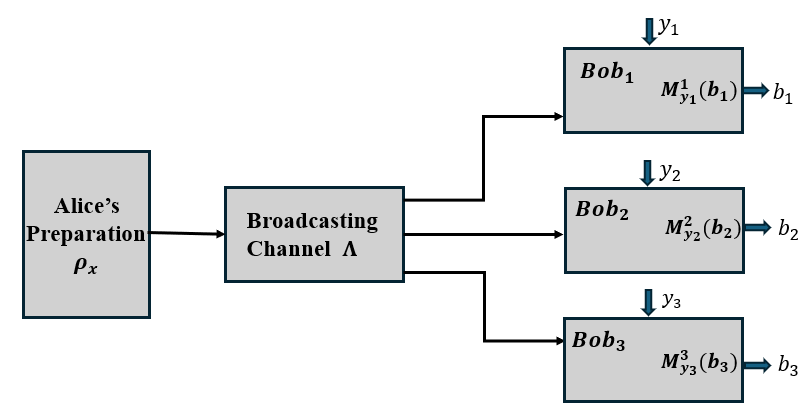} }}%
		\caption{Schematic diagram for sharing quantum advantage to multiple observers by (a) sequential sharing strategy, (b) parallel sharing strategy.}%
		\label{fig:sharing}%
	\end{figure}

In this paper, we are interested in the scenario where multiple observers perform prepare-measure tasks. In general, we can consider one preparer and a multiple number of independent receivers. One can devise such scenarios in the following two ways as also shown in Fig.~\ref{fig:sharing}. In both cases, the system sent by Alice to the Bobs is a $d$-dimensional quantum system, so that the two strategies can be compared on equal footing.

\emph{\textbf{Sequential sharing:-}}
In the sequential task, Alice prepares the input quantum state by performing a measurement $N_x=\{N_x(a)\}_a$ on her share of a bipartite state $\rho_{AB}\in\cS(\cH_A\otimes\cH_B)$, depending on the classical input $x$. She then sends system $B$ to the first observer, Bob$_1$. After receiving the system, Bob$_1$ applies an instrument $\cI^{y_1}=\{\Phi^{y_1}_{b_1}\}_{b_1}\subset\Inst(\cH_B,\cH_B)$ depending on his input $y_1$, obtains the outcome $b_1$, and forwards the post-measurement system to Bob$_2$, who does the same. The chain continues until the last observer, Bob$_n$, who simply performs a measurement $M^{n}_{y_n}=\{M^{n}_{y_n}(b_n)\}_{b_n}\in\Meas(\cH_B)$. Here each $\Phi^{y}_{b}$ is a completely positive trace-nonincreasing map, and $\sum_{b}\Phi^{y}_{b}$ is a quantum channel for every $y$. We denote the resulting probability distribution by $p_{\mathrm{seq}}(a,\Vec{b}\,|\,x,\Vec{y})$, where $\Vec{b}=(b_1,\dots,b_n)$ and $\Vec{y}=(y_1,\dots,y_n)$. Since the instruments act one after another, they are composed rather than tensored, and the distribution reads
\bea\label{prob_seq}
p_{\mathrm{seq}}(a,\Vec{b}\,|\,x,\Vec{y})
&=&\tr\Big[\big(\idch_A\otimes\Phi^{y_{n-1}}_{b_{n-1}}\circ\cdots\circ\Phi^{y_1}_{b_1}\big)(\rho_{AB})\nonumber\\
&&\hspace{1.5cm}\times\big(N_x(a)\otimes M^{n}_{y_{n}}(b_{n})\big)\Big],
\eea
where $M^{n}_{y_n}(b_n)$ is the measurement implemented by the last Bob. The preceding Bobs ($i<n$) act through instruments, so their outcomes $b_i$ enter the distribution through the maps $\Phi^{y_i}_{b_i}$. In particular, the choice $y_1$ can influence the statistics observed by Bob$_2,\dots,$Bob$_n$.

\emph{\textbf{Parallel sharing:-}}
The preparation procedure is the same as in the sequential strategy. After the preparation, Alice applies a channel $\Lambda\in\Chan(\cH_B,\cH_{B_1}\otimes\cH_{B_2}\otimes\cdots\otimes\cH_{B_n})$ and sends system $B_i$ to Bob$_i$, who measures $M^i_{y_i}\in\Meas(\cH_{B_i})$. We place no restriction on $\Lambda$, but we require $\dim\cH_{B_i}=d$ for all $i$, i.e.\ every party receives a system of the same dimension $d$, equal to that of the system in the sequential scenario. The probability distribution takes the form
\bea\label{prob_par}
&&p_{\mathrm{par}}(a,\Vec{b}\,|\,x,\Vec{y})\nonumber\\
&=&\tr\Big[(\idch_A\otimes\Lambda)(\rho_{AB})\,
\Big(N_x(a)\otimes\bigotimes_{i=1}^n M^{i}_{y_i}(b_i)\Big)\Big].
\eea
It is important to note that, within the framework of a parallel strategy, the marginal statistics of any subset of observers can be obtained from the joint probability distribution given in Eq.~\eqref{prob_par}. This is formally established in the following.
\begin{lemma}\label{Lem:NS_par}
Let $p_{\mathrm{par}}(a,\vec b\,|\,x,\vec y)$ be any distribution of the form \eqref{prob_par}. For every subset $S\subseteq\{1,\dots,n\}$ with complement $T=\{1,\dots,n\}\setminus S$, the marginal
\begin{equation}
\label{eq:bob_n_independent}
p_{\mathrm{par}}(\vec b_T\,|\,x,\vec y):=\sum_{a}\sum_{\vec b_S}p_{\mathrm{par}}(a,\vec b\,|\,x,\vec y)
\end{equation}
is independent of $x$ and of $\vec y_S$. This implies that the marginal $\sum_{a,\vec b_S}p_{\mathrm{par}}(a,\vec b\,|\,x,\vec y)$, is independent of $\vec y_S$. In particular, for every $i$,
\begin{equation}\label{eq:NS_n}
p_{\mathrm{par}}(\vec b_{\neq i}\,|\,x,y_i,\vec y_{\neq i})=p_{\mathrm{par}}(\vec b_{\neq i}\,|\,x,y_i',\vec y_{\neq i})\qquad\text{for all } y_i,y_i'.
\end{equation}
\end{lemma}
\begin{proof}
The post-interaction state after the application of the channel $\Lambda$ is defined as $\tilde\rho:=(\idch_A\otimes\Lambda)(\rho_{AB})$, which lives on
$\cH_A\otimes\cH_{B_1}\otimes\cdots\otimes\cH_{B_n}$. Now, summing over $a$ and over all $b_i$ with $i\in S$ of Eq.~\eqref{prob_par}, and using
$\sum_a N_x(a)=\Id_A$, and $\sum_{b_i}M^i_{y_i}(b_i)=\Id_{B_i}$ for each $i\in S$, we get
\begin{align}
\sum_{a}\sum_{\vec b_S}p_{\mathrm{par}}(a,\vec b\,|\,x,\vec y)
&=\tr\Big[\tilde\rho\,\Big(\Id_A\otimes\bigotimes_{i\in S}\Id_{B_i}\otimes\bigotimes_{j\in T}M^{j}_{y_j}(b_j)\Big)\Big]\nonumber\\
&=\tr\Big[\Lambda(\rho_B)\,\Big(\bigotimes_{i\in S}\Id_{B_i}\otimes\bigotimes_{j\in T}M^{j}_{y_j}(b_j)\Big)\Big],
\end{align}
where $\rho_B=\tr_A[\rho_{AB}]$. In the second line we used
$\tr[(\idch_A\otimes\Lambda)(\rho_{AB})(\Id_A\otimes X)]=\tr[\Lambda(\rho_B)X]$, which holds because $\tr_A\big[(\idch_A\otimes\Lambda)(\rho_{AB})\big]=\Lambda(\tr_A\rho_{AB})$. The right-hand side depends only on $\Lambda$, $\rho_B$ and $(y_j)_{j\in T}$, so it is independent of $x$ and of $\vec y_S$. Taking $S=\{i\}$ gives \eqref{eq:bob_n_independent}.

To establish the second claim, for any no-signaling distribution, summing over $\vec b_S$ yields,
\begin{equation}
\sum_{a,\vec b_S}p_{\mathrm{par}}(a,\vec b\,|\,x,\vec y)
=\tr\Big[\tilde\rho\,\Big(N_x(a)\otimes\bigotimes_{i\in S}\Id_{B_i}\otimes\bigotimes_{j\in T}M^{j}_{y_j}(b_j)\Big)\Big],
\end{equation}
which again does not depend on $\vec y_S$.
\end{proof}
\begin{remark}
Lemma~\ref{Lem:NS_par} does not hold for the sequential strategy characterized by  Eq.~\eqref{prob_seq}. In this case, the instrument of Bob$_1$ acts on the system subsequently received by Bob$_2$, and so on. Consequently, each preceding measurement can alter the marginal statistics of any subsequent observer. This constitutes the essential operational difference between the two strategies.
\end{remark}
\section{Comparing parallel and Sequential statistics}\label{sec:compare}
We begin by investigating the hierarchy between sequential and parallel sharing of nonclassicality. We consider the simplest scenario involving two observers and ask whether nonclassical statistics that can be shared between the two such observers in one strategy can also be shared in the other, and vice versa.
\begin{proposition}
\label{Prop:par_to_seq}
Every probability distribution $\{p(a,b_1,b_2\,|\,x,y_1,y_2)\}$ achievable through a parallel strategy can also be reproduced by a sequential strategy, but the converse does not hold.
\end{proposition}
\begin{proof}
The probability distributions in sequential and parallel strategy are given in \eqref{prob_seq} and \eqref{prob_par} respectively. Then for a scenario involving two observers, Bob$_1$ and Bob$_2$, the outcome statistics corresponding to the two strategies can be written, respectively, as follows,
\begin{align}
    p_{\mathrm{par}}&(a,b_1,b_2\,|\,x,y_1,y_2)
    =\tr\Big[(\idch_A\otimes\Lambda)(\rho_{AB})\nonumber\\
    &\times\big(N_x(a)\otimes M^{1}_{y_1}(b_1)\otimes M^{2}_{y_2}(b_2)\big)\Big].
    \label{Eq:para_prob}
\end{align}
and
\begin{align}
    p_{\mathrm{seq}}&(a,b_1,b_2\,|\,x,y_1,y_2)
    =\tr\Big[(\idch_A\otimes\Phi^{y_1}_{b_1})(\rho_{AB})\nonumber\\
    &\times\big(N_x(a)\otimes M^{2}_{y_2}(b_2)\big)\Big].
    \label{Eq:seq_prob}
\end{align}
Now, we construct, for every $y_1$, an instrument $\cI^{y_1}=\{\Phi^{y_1}_{b_1}\}_{b_1}\in\Inst(\cH_B,\cH_B)$ for sequential strategy such that $p_{\mathrm{seq}}=p_{\mathrm{par}}$ for all $a,b_1,b_2,x,y_1,y_2$. For $\sigma\in\cL(\cH_B)$ and $b_1\in\Omega_{M^1_{y_1}}$, we define
\begin{align}
    \Phi^{y_1}_{b_1}(\sigma):=\tr_{B_1}\Big[&\Big(\sqrt{M^{1}_{y_1}(b_1)}\otimes\Id_{B_2}\Big)\,\Lambda(\sigma)\nonumber\\
    &\times\Big(\sqrt{M^{1}_{y_1}(b_1)}\otimes\Id_{B_2}\Big)\Big].
    \label{Eq:para_in_convert}
\end{align}
Each $\Phi^{y_1}_{b_1}$ is completely positive, as it is a composition of the channel $\Lambda$, a conjugation by $\sqrt{M^{1}_{y_1}(b_1)}\otimes\Id_{B_2}$, and the partial trace. Using the cyclicity of the partial trace over $B_1$ and $\sum_{b_1}M^1_{y_1}(b_1)=\Id_{B_1}$,
\begin{equation}
    \sum_{b_1}\Phi^{y_1}_{b_1}(\sigma)=\tr_{B_1}\big[\Lambda(\sigma)\big],
\end{equation}
which is a channel in $\Chan(\cH_B,\cH_{B_2})$. Hence $\cI^{y_1}\in\Inst(\cH_B,\cH_B)$.
Now, substituting Eq.~\eqref{Eq:para_in_convert} into Eq.~\eqref{Eq:seq_prob} and using the cyclicity of the trace we obtain, 
\begin{align}
&p_{\mathrm{seq}}(a,b_1,b_2\mid x,y_1,y_2)
\nonumber\\
&=\operatorname{Tr}_{AB_2}\Big[
(\operatorname{id}_A\otimes\Phi_{b_1}^{y_1})(\rho_{AB})
\big(N_x(a)\otimes M_{y_2}^{2}(b_2)\big)
\Big]
\nonumber\\[6pt]
&=\operatorname{Tr}_{AB_2}\Bigg[
\operatorname{Tr}_{B_1}\Big[
\big(I_A\otimes\sqrt{M_{y_1}^{1}(b_1)}
\otimes I_{B_2}\big)
(\operatorname{id}_A\otimes\Lambda)(\rho_{AB})
\nonumber\\
&\qquad \ \ \ \ \ \times
\big(I_A\otimes\sqrt{M_{y_1}^{1}(b_1)}
\otimes I_{B_2}\big)
\Big]
\big(N_x(a)\otimes M_{y_2}^{2}(b_2)\big)
\Bigg]
\nonumber\\[6pt]
%&=\operatorname{Tr}_{AB_1B_2}\Big[
%\big(I_A\otimes\sqrt{M_{y_1}^{1}(b_1)}
%\otimes I_{B_2}\big)
%(\operatorname{id}_A\otimes\Lambda)(\rho_{AB})
%\nonumber\\
%&\qquad \times
%\big(I_A\otimes\sqrt{M_{y_1}^{1}(b_1)}
%\otimes I_{B_2}\big)
%\big(N_x(a)\otimes I_{B_1}\otimes M_{y_2}^{2}(b_2)\big)
%\Big]
%\nonumber\\[6pt]
&=\operatorname{Tr}\Big[
(\operatorname{id}_A\otimes\Lambda)(\rho_{AB})
\big(I_A\otimes\sqrt{M_{y_1}^{1}(b_1)}
\otimes I_{B_2}\big)
\nonumber\\
&\qquad\times
\big(N_x(a)\otimes I_{B_1}\otimes M_{y_2}^{2}(b_2)\big)
\big(I_A\otimes\sqrt{M_{y_1}^{1}(b_1)}
\otimes I_{B_2}\big)
\Big]
\nonumber\\[6pt]
&=\operatorname{Tr}\Big[
(\operatorname{id}_A\otimes\Lambda)(\rho_{AB})
\big(N_x(a)\otimes
M_{y_1}^{1}(b_1)\otimes M_{y_2}^{2}(b_2)\big)
\Big]
\nonumber\\[6pt]
&=p_{\mathrm{par}}(a,b_1,b_2\mid x,y_1,y_2).
\end{align}
which proves the first part of the proposition.
For the second part, we provide existence of a sequential strategy which violates lemma \ref{Lem:NS_par}.
Take $\cH_B=\mathbb{C}^2$ with $\rho_{AB}=\rho_A\otimes|0\rangle\langle0|$. Let Alice have a single input and a single outcome, so $N_x(a)=\Id_A$. Let Bob$_1$ have inputs $y_1\in\{0,1\}$ and outcomes $b_1\in\{0,1\}$. We define the map that characterizes the instrument implementing this measurement as
\begin{equation}
\Phi^{y_1}_{b_1}(\sigma):=\tfrac12\tr(\sigma)|y_1\rangle\langle y_1| ,\qquad \forall b_1\in\{0,1\}.
\end{equation}
Each $\Phi^{y_1}_{b_1}$ is completely positive, and
\begin{equation}
\sum_{b_1}\Phi^{y_1}_{b_1}(\sigma)=\tr(\sigma)|y_1\rangle\langle y_1|
\end{equation}
is a channel (measure-and-prepare), so $\cI^{y_1}=\{\Phi^{y_1}_{b_1}\}_{b_1}$ is a valid instrument. Bob$_2$ has a single input and measures in the computational basis: $M^2_{y_2}(b_2)=\{|b_2\rangle\langle b_2|\}_{b_2}$. Then by Eq. \eqref{Eq:seq_prob} we have,
\begin{align}
&p_{\mathrm{seq}}(a,b_1,b_2|x,y_1,y_2)\nonumber\\
&=\tr\Big[(\idch_A\otimes\Phi^{y_1}_{b_1})(\rho_A\otimes|0\rangle\langle0|)\big(N_x(a)\otimes M^2_{y_2}(b_2)\big)\Big]\nonumber\\
&=\tr\big[\rho_A,N_x(a)\big]\tr\Big[\Phi^{y_1}_{b_1}\big(|0\rangle\langle0|\big)|b_2\rangle\langle b_2|\Big]\nonumber\\
&=\tfrac12\delta_{b_2,y_1}.
\end{align}
Summing over $a$ and $b_1$ gives the marginal of Bob$_2$,
\begin{equation}
p_{\mathrm{seq}}(b_2|,x,y_1,y_2)=\sum_{a,b_1}p_{\mathrm{seq}}(a,b_1,b_2,|,x,y_1,y_2)=\delta_{b_2,y_1}.
\end{equation}

This depends on $y_1$, such that for $y_1=0$, Bob$_2$ always obtains $b_2=0$, whereas for $y_1=1$, he always obtains $b_2=1$. So, Bob$_1$’s input is perfectly transmitted to Bob$_2$. However, for the current scenario with parallel distribution no-signaling condition \eqref{eq:NS_n} gives, 
\begin{equation}
p_{\mathrm{par}}(b_2,|,x,y_1,y_2)=p_{\mathrm{par}}(b_2,|,x,y_1',y_2)\quad\text{for all }y_1,y_1'.
\end{equation}
This contradicts with the sequential statistics.
 Hence the set of distributions obtainable with the parallel strategy is a strict subset of those obtainable with the sequential strategy. Together with the inclusion proved earlier, this establishes the proposition.
\end{proof}

At this point, a natural question arises: What happens when we consider more than two observers? From the last proposition, it is straight-forward to prove that every probability distribution obtainable with the parallel strategy can also be obtained with the sequential strategy, with forwarded systems
\begin{equation}   \cH_{S_{k}}=\cH_{B_k}\otimes\cH_{B_{k+1}}\otimes\cdots\otimes\cH_{B_N},\qquad k=2,\dots,N.
    \label{Eq:forwarded}
\end{equation}
In order to simulate
the statistics of parallel strategies, the sequential strategy requires Bob$_k$ to forward a system of dimension $d^{N-k}$. We now show that if each Bob forwards a system of dimension $d$ to the subsequent observer, the statistics of the parallel strategy cannot, in general, be reproduced. 

We work in the prepare-and-measure version of the scenario, and keep Alice's preparations fixed. In the sequential strategy, Bob$_k$ applies instrument $\cI^{k,y_k}_\lambda=\{\Phi^{k,y_k}_{b_k,\lambda}\}_{b_k}\in\Inst(\cH_B,\cH_B)$, where $\lambda$ is classical randomness shared by all Bobs and distributed with $\pi(\lambda)$. Shared randomness can only help the sequential strategy. Further, let $d=2$ and $N=3$, with Alice preparing states from the set $\rho_x\in\{\ket{0}\bra{0},\ket{1}\bra{1},\ket{+}\bra{+}, \ket{-}\bra{-}\}$. Consider the broadcasting channel $\Lambda\in\Bcast_{1\rightarrow3}(\mathbb{C}^2)$ whose action on the state $\rho$ is given by,
\be\label{Eq:counter_channel}
\Lambda(\rho)=\ket{\Phi^+}\bra{\Phi^+}_{B_1B_3}\otimes\rho_{B_2}.
\ee
This sends Alice's qubit to Bob$_2$ and subsystems from maximally entangled state $\ket{\Phi^+}\bra{\Phi^+}_{B_1B_3}$ to Bob$_1$ and Bob$_3$. Subsequently, Bob$_1$ measures $A_0=\sigma_z$ or $A_1=\sigma_x$, Bob$_3$ measures $B_{0,1}=(\sigma_z\pm\sigma_x)/\sqrt2$, and Bob$_2$ measures $\sigma_z$ or $\sigma_x$. The resulting distribution factorizes as
\begin{equation}
   \label{Eq:counter_target}
    p_{\mathrm{par}}(b_1,b_2,b_3\,|\,x,\vb y)=p_{\mathrm{T}}(b_1,b_3\,|\,y_1,y_3)\,\tr\big[\rho_x\,\Pi^{y_2}_{b_2}\big],
\end{equation}
where $\{\Pi^{0}_{b}\}$ and $\{\Pi^{1}_{b}\}$ are the eigenprojectors of $\sigma_z$ and $\sigma_x$ respectively.Since the state shared between Bob$_1$ and Bob$_3$ is maximally entangled, it follows from Eq.~\eqref{Eq:counter_target} that the correlations described by $\{p_{\mathrm{T}}(b_1,b_3\,|\,y_1,y_3)\}_{b_1,b_3,y_1,y_3}$ attain Tsirelson's bound for the Bell-CHSH expression, i.e., $\mathcal{B}=\sum_{y_1,y_3}(-1)^{y_1y_3}\langle b_1b_3\rangle=2\sqrt2$~\cite{chsh69,Cirelson1980}, with $\langle b_1b_3\rangle:=\sum_{b_1,b_3}(-1)^{b_1+b_3}p_{\mathrm T}(b_1,b_3\,|\,y_1,y_3)$ and $\mathcal{B}$ being the Bell operator that appears in Bell-CHSH inequality. We now prove that the probability distribution given in Eq.~\eqref{Eq:counter_target} cannot be reproduced by any sequential strategy in which the observers successively share a single quantum system of the same dimension as the initially prepared system. This requires to first prove the following results.

\begin{lemma}
\label{lem:projectors_topovm}
Let $\Theta:\mathcal L(\mathbb C^2)\to\mathcal L(\mathbb C^2)$
be a quantum channel, and let
$Q_0=\{Q_0(b)\}$ and $Q_1=\{Q_1(b)\}$ be two binary POVMs.
Suppose the effective measurements $\Theta^\dagger(Q_0)$
and $\Theta^\dagger(Q_1)$ perfectly distinguish the
orthogonal pairs $\{|0\rangle,|1\rangle\}$ and
$\{|+\rangle,|-\rangle\}$, respectively. Then
\begin{align}
\Theta^\dagger(Q_0(0))=|0\rangle\langle0|,
\qquad
\Theta^\dagger(Q_1(0))=|+\rangle\langle+|.
\end{align}
\end{lemma}

\begin{proof}
Since $\Theta$ is a CPTP map,
its adjoint $\Theta^\dagger$ is unital and completely positive.
Consequently, we may write the effective operator as
$F=\Theta^\dagger(Q_0(0))$ which satisfies $0\leq F\leq I$. Now, the perfect discrimination of $|0\rangle$ and $|1\rangle$ requires
\begin{align}
\langle0|F|0\rangle=1,
\qquad
\langle1|F|1\rangle=0.
\end{align}
Without loss of generality we can write $F$ in the computational basis as,
\begin{align}
F=
\begin{pmatrix}
1 & c\\
c^* & 0
\end{pmatrix},
\end{align}
positivity implies
$\det F=-|c|^2\geq0$, and hence $c=0$. Therefore we must have,
\begin{align}
\label{eq:proj1}
\Theta^\dagger(Q_0(0))=|0\rangle\langle0|.
\end{align}
Applying the same argument in the basis
$\{|+\rangle,|-\rangle\}$ gives
\begin{align}
\label{eq:proj2}
\Theta^\dagger(Q_1(0))=|+\rangle\langle+|.
\end{align}
This completes the proof.
\end{proof}

\begin{lemma}
\label{lem:map_unitary}
Let $\Theta:\mathcal L(\mathbb C^2)\to\mathcal L(\mathbb C^2)$ be a quantum channel such that $\Theta^\dagger(Q_0(0))=|0\rangle\langle0|$ and $\Theta^\dagger(Q_1(0))=|+\rangle\langle+|$, where $Q_0(0)$ and $Q_1(0)$ are two POVM effects. Then $\Theta$ is necessarily a unitary channel, i.e., $\Theta(\rho)=U\rho U^\dagger$ for some unitary $U\in\mathcal L(\mathbb C^2)$.
\end{lemma}

\begin{proof}
Since $\Theta^\dagger$ is completely positive and unital,
it satisfies the Kadison--Schwarz (KS) inequality,
\begin{align}
\Theta^\dagger(Q^2)\geq
\left[\Theta^\dagger(Q)\right]^2,
\end{align}
for every positive operator $Q$.Although complete positivity is not necessary for the inequality to hold, every $2$-positive map satisfies the KS inequality. Now for the operator
$Q\in\{Q_0(0),Q_1(0)\}$, if $\Theta^\dagger(Q)=P$, where $P$ is a
rank-one projector then, $\Theta^\dagger(Q^2)\geq P^2=P$. On the other hand, $0\leq Q\leq I$ implies $Q^2\leq Q$. Then by positivity, $\Theta^\dagger(Q^2)
\leq\Theta^\dagger(Q)=P$. Combination of these two aforementioned inequalities yields
\begin{align}
\Theta^\dagger(Q^2)=
\left[\Theta^\dagger(Q)\right]^2.
\end{align}

Consequently, both $Q_0(0)$ and $Q_1(0)$ belong to the
multiplicative domain $\mathcal D$ of $\Theta^\dagger$,
which is a $C^*$-subalgebra on which $\Theta^\dagger$
acts as a $*$-homomorphism.

Now, the image $\Theta^\dagger(\mathcal D)$ contains
the two noncommuting projectors $ P_Z=|0\rangle\langle0|$, $P_X=|+\rangle\langle+|$.
These projectors generate the full matrix algebra
$\mathcal L(\mathbb C^2)$, since
\begin{align}
\sigma_z=2P_Z-I,\qquad
\sigma_x=2P_X-I,\qquad
\sigma_y=-i\sigma_z\sigma_x.
\end{align}
As $\Theta^\dagger(\mathcal D)$ is itself a unital
$*$-subalgebra, it follows that
\begin{align}
\Theta^\dagger(\mathcal D)=\mathcal L(\mathbb C^2).
\end{align}
Therefore, $\dim\mathcal D\geq4$.
Since $\mathcal D\subseteq\mathcal L(\mathbb C^2)$,
we conclude that
$\mathcal D=\mathcal L(\mathbb C^2)$.
Thus, $\Theta^\dagger$ is a surjective unital
$*$-homomorphism of the full matrix algebra and hence
a $*$-automorphism. Every $*$-automorphism of $\mathcal L(\mathbb C^2)$
is implemented by a unitary operator \cite{Paulsen_2003}. Consequently,
$\Theta^\dagger(X)=U^\dagger XU$,
which further implies $\Theta(\rho)=U\rho U^\dagger$.
This completes the proof.
\end{proof}

\begin{theorem}\label{Thm:no_dim_d}
No sequential strategy involving three observers acting successively on a fixed, initially prepared quantum system, with each observer forwarding a system of dimension $d=2$ to the next, can reproduce the distribution~\eqref{Eq:counter_target}, even with arbitrary shared randomness among the observers.
\end{theorem}

\begin{proof}
We prove the theorem by contradiction. To this end, we first assume that such a sequential strategy exists. To establish the result in full generality, we allow the sequential observers to share an arbitrary random variable $\lambda$, distributed according to a probability density $\pi(\lambda)>0$. We drop $\lambda$ from the notation when no confusion arises. Let $\Theta_{y_1}:=\sum_{b_1}\Phi^{1,y_1}_{b_1}$ be Bob$_1$'s average channel, and let $Q_{y_2}(b_2):=\big(\Phi^{2,y_2}_{b_2}\big)^\dagger(\Id)$ be the measurement on the forwarded qubit induced by Bob$_2$'s instrument. It does not depend on the measurement $y_1$ or the outcome $b_{1}$, since Bob$_2$ does not know any of the Bob$_1$'s actions or its outcomes.

The Kraus operators of all $\Phi^{1,y_1}_{b_1}$ together form a Kraus decomposition of $\Theta_{y_1}$. Since the Choi operator of a unitary channel has rank one, every such Kraus operator is proportional to $U_{y_1}$. Hence
\begin{equation}
    \Phi^{1,y_1}_{b_1}(\rho)=p(b_1\,|\,y_1)\,U_{y_1}\rho\,U_{y_1}^\dagger,
\end{equation}
with probabilities $p(b_1\,|\,y_1)$ that do not depend on $\rho$. Now, from Lemma \ref{lem:projectors_topovm} (acting $\Theta$ from left to both side of Eq. \eqref{eq:proj1}), and \eqref{eq:proj2}), and using Lemma \ref{lem:map_unitary}, we may write $\Theta(\ket0\bra0)=U\ket0\bra0 U^\dagger= Q_0(0)$ and $\Theta(\ket+\bra+)=U_{y_1}\ket+\bra+U_{y_1}^\dagger=Q_1(0)$ for both $y_1$. Thus $V:=U_1^\dagger U_0$ commutes with $\ket0\bra0$ and $\ket+\bra+$, and hence with all of $\cL(\mathbb{C}^2)$, so $V$ is just a phase. Therefore the forwarded state $U\rho_xU^\dagger$ by Bob$_{1}$ is independent of $y_1$ and $b_1$.

 It follows that, after marginalizing over Bob$_2$'s outcome, the correlations between Bob$_1$ and Bob$_3$ necessarily factorize in terms of the shared random variable $\lambda$ as,
\begin{equation}
    p_{C}(b_1,b_3\,|\,x,\vec y)=\sum_\lambda\pi(\lambda)\,p(b_1\,|\,y_1,\lambda)\,q(b_3\,|\,x,y_2,y_3,\lambda),
\end{equation} where the probability $q$ contains everything that happens after Bob$_1$. For fixed $x$ and $y_2$, this is a local hidden-variable model for the pair $(y_1,y_3)$, and therefore with the probability distribution $\{ p_{C}(b_1,b_3\,|\,x,\vec y)\}_{b_{1},b_{2},y_{1},y_{2}}$ the  Bell-CHSH value should be bounded as $\mathcal{B}\leq 2$~\cite{chsh69}. This contradicts the maximum attainable value of the Bell-CHSH expression calculated from the marginal probabilities of the parallel strategy given in Eq.~\eqref{Eq:counter_target}.
\end{proof}

\section{Parallel sharing of nonclassical correlations in communication tasks with linear figure of merit}\label{Sec:two_setting}

Consider a multiparty prepare-and-measure communication task involving $Y=n$ measurement settings for the decoding process. We further assume that the success of the task is characterized by a linear figure of merit given by
\begin{equation}\label{Eq:gen_merit}
\cW[p]=\sum_{x,a}\sum_{y\in[n]}\sum_{b}c(x,y,a,b)\,p(a,b|x,y),
\end{equation}
with arbitrary real coefficients $c(x,y,a,b)$. For an assemblage $\{\rho_{a|x}\}$ and measurements $M_1,\dots,M_n\in\Meas(\cH_B)$ we write $\cW(M_1,\dots,M_n):=\cW[p]$, with $p(a,b|x,y)=\tr[\rho_{a|x}M_y(b)]$. Then
\bea\label{Eq:additivity}
&&\cW(M_1,\dots,M_n)=\sum_{y\in[n]}V_y(M_y),\qquad\\
&& \text{with, \ }V_y(M_y):=\sum_{x,a,b}c(x,y,a,b)\,\tr\big[\rho_{a|x}M_y(b)\big].
\eea
For two measurement settings $s$ and $t$, and with $s\ne t$ we define the two-setting merit as
\begin{equation}\label{Eq:pair_merit}
\cW^{st}(M_s,M_t):=V_s(M_s)+V_t(M_t).
\end{equation}

The \emph{classical bound} $C_d(\cW)$ is the maximum of $\cW[p]$ over all $p$ of the form \eqref{c_stat}, allowing in addition shared randomness $\lambda$ with $p_e(a,m|x,\lambda)$ and $p_d(b|m,y,\lambda)$. Since $\cW$ is linear, shared randomness does not increase the maximum, so it can be attained by deterministic strategies. The classical figure of merit for two measurement settings $s,t$, is defined as $C_d(\cW^{st})$ in the same way.Moreover, we define the average of the pairwise classical figures of merit for arbitrary number of measurement settings as follows:
\begin{equation}\label{Eq:Ctilde}
\widetilde C_d(\cW):=\frac{1}{n-1}\sum_{s<t}C_d(\cW^{st}).
\end{equation}
\begin{comment}
 \begin{lemma}\label{Lem:Ctilde}
For $n\ge2$, $C_d(\cW)\le\widetilde C_d(\cW)$, with equality for $n=2$.
\end{lemma}
\begin{proof}
In the expression for the average figure of merit, each $V_y$, except for $V_1$ and $V_n$, appears exactly twice, yielding $n-1$ pairs $\{s,t\}$ of terms in the sum. Therefore, $\cW=\frac1{n-
1}\sum_{s<t}\cW^{st}$ holds for every strategy. Now, maximizing a sum is at most the sum of the maxima, which gives the inequality. For $n=2$ there is only one pair and $\widetilde C_d(\cW)=C_d(\cW)$.
\end{proof}   
\end{comment}
\begin{lemma}\label{Lem:Ctilde}
For $n\geq 2$, the classical bound $C_d(W)$ and the average
pairwise classical bound $\widetilde C_d(W)$ satisfy $C_d(W)\leq \widetilde C_d(W)$. Moreover, equality holds for $n=2$.
\end{lemma}

\begin{proof}
Consider the sum of $\cW^{st}$ in Eq. \eqref{Eq:additivity} over all distinct pairs
$s<t$. Since each measurement setting $y\in[n]$ appears
in exactly $n-1$ distinct pairs, every term $V_y$
contributes exactly $n-1$ times to the sum. Therefore,
for any given strategy,
\begin{align}
    \sum_{s<t}\cW^{st}
    &=\sum_{s<t}(V_s+V_t)=(n-1)\sum_{y=1}^{n}V_y\nonumber =(n-1)\cW.
\end{align}
Consequently we have,
\begin{align}\label{Eq:AverageMerit}
    \cW=\frac{1}{n-1}\sum_{s<t}\cW^{st}.
\end{align}

Let $ \mathcal C_d$ denote the set of all classical
prepare-and-measure strategies employing a $d$-level
classical message, including strategies assisted by
shared randomness. By definition, the classical bound
of $\cW$ is the maximum value attained over $ \mathcal C_d$.
Thus,
\begin{align}
    C_d(\cW)
    &=\max_{\mathcal S\in\mathcal C_d}\cW(\mathcal S)\nonumber\\
    &=\max_{\mathcal S\in\mathcal C_d}
    \left[
    \frac{1}{n-1}
    \sum_{s<t}\cW^{st}(\mathcal S)
    \right]\nonumber\\
    &\leq\frac{1}{n-1}
    \sum_{s<t}
    \max_{\mathcal S\in \mathcal C_d}
    \cW^{st}(\mathcal S)\nonumber\\
    &=\frac{1}{n-1}
    \sum_{s<t}C_d(\cW^{st})=\widetilde C_d(\cW).
\end{align}
Here, the inequality follows from the fact that the
maximum of a sum of functions over a common set
cannot exceed the sum of their individual maxima.

Finally, for $n=2$, there is only one pair of
measurement settings, namely $(s,t)=(1,2)$.
Therefore, $\cW=\cW^{12}$ and
\begin{align}
    \widetilde C_d(W)
    =C_d(W^{12})
    =C_d(W).
\end{align}
This completes the proof.
\end{proof}

\begin{lemma}\label{Lem:cross}
For any two distinct observers $k\neq l$ and arbitrary
measurement settings $y,y'\in[n]$, the effective
measurements $E_y^k$ and $E_{y'}^l$ are jointly measurable.
\end{lemma}

\begin{proof}
Consider the effective measurements
$E_y^k=\{E_y^k(b)\}_b$ and
$E_{y'}^l=\{E_{y'}^l(b')\}_{b'}$, induced by the
broadcasting channel $\Lambda$. We construct a joint
measurement $G=\{G(b,b')\}_{b,b'}$ on the input
Hilbert space as
\begin{align}
G(b,b')
:=\Lambda^\dagger\left(
\cdots\otimes F_y^k(b)\otimes\cdots
\otimes F_{y'}^l(b')\otimes\cdots
\right),
\end{align}
where $F_y$ are local POVM elements and the identity operators act on all remaining output
subsystems.
Since $M_y^k(b)$ and $M_{y'}^l(b')$ are positive
semidefinite and $\Lambda^\dagger$ is completely
positive, we have $G(b,b')\geq0$. Moreover, using
the normalization of the local POVMs and the
unitality of $\Lambda^\dagger$, we obtain,
$\sum_{b,b'}G(b,b')
=\Lambda^\dagger(\mathbb{1}_{\mathrm{out}})
=\mathbb{1}_{B}$.
Thus, $G$ is a valid POVM. Finally, taking the marginals gives
\begin{align}
\sum_{b'}G(b,b')&=E_y^k(b), \ 
\sum_bG(b,b')=E_{y'}^l(b').
\end{align}
Hence, $G$ as a joint POVM for $E_y^k$ and
$E_{y'}^l$, establishing their joint measurability.
\end{proof}
\begin{lemma}[Joint measurability implies classical simulability]\label{Lem:JM_classical}
Let $\{\rho_{a|x}\}$ be an arbitrary assemblage on
$\mathbb{C}^d$. If two measurements
$E_s,E_t\in\mathcal M(\mathbb{C}^d)$ are jointly
measurable, then
\begin{align}
    W^{st}(E_s,E_t)\leq C_d(W^{st}).
\end{align}
\end{lemma}

\begin{proof}
Since $E_s$ and $E_t$ are jointly measurable, there
exists a parent POVM $G=\{G(g)\}_g$ and conditional
probability distributions $p_y(b|g)$ such that
\begin{align}
    E_y(b)=\sum_g p_y(b|g)G(g),
    \qquad y\in\{s,t\}.
\end{align}
Consequently, the corresponding quantum statistics
can be expressed as
\begin{align}
    p(a,b|x,y)
    &=\operatorname{Tr}[\rho_{a|x}E_y(b)]
    \nonumber\\
    &=p(a|x)\sum_g p_y(b|g)
    \operatorname{Tr}[\hat{\rho}_{a|x}G(g)],
    \label{eq:joint_classical_sim}
\end{align}
where $p(a|x)=\operatorname{Tr}[\rho_{a|x}]$ and
$\hat{\rho}_{a|x}=\rho_{a|x}/p(a|x)$ whenever
$p(a|x)>0$. Terms with $p(a|x)=0$ contribute
nothing to the statistics.

We now invoke the result of Frenkel and Weiner
\cite{fw}, which establishes that the
statistics of any single POVM on a finite family
of $d$-dimensional quantum states can be reproduced
using a classical $d$-level message assisted by
shared randomness. Applying this result to the
states $\{\hat{\rho}_{a|x}\}$ and the POVM $G$,
there exist shared randomness $\lambda$ distributed
according to $\pi(\lambda)$, classical encoding
probabilities $p(m|x,a,\lambda)$ with $m\in[d]$,
and decoding probabilities $p(g|m,\lambda)$ such that
\begin{align}
    \operatorname{Tr}[\hat{\rho}_{a|x}G(g)]
    =\sum_{\lambda,m}\pi(\lambda)
    p(m|x,a,\lambda)p(g|m,\lambda).
\end{align}

Now the classical encoding and decoding can be defined as
distributions as
\begin{align}
    p_{\mathrm e}(a,m|x,\lambda)
    &:=p(a|x)p(m|x,a,\lambda),\\
    p_{\mathrm d}(b|m,y,\lambda)
    &:=\sum_g p(g|m,\lambda)p_y(b|g).
\end{align}
Both distributions are normalized, since
\begin{align}
    \sum_{a,m}p_{\mathrm e}(a,m|x,\lambda)
    &=\sum_a p(a|x)=1,\\
    \sum_b p_{\mathrm d}(b|m,y,\lambda)
    &=\sum_g p(g|m,\lambda)=1.
\end{align}
Substituting these expressions into
Eq.~\eqref{eq:joint_classical_sim}, we obtain for $y\in\{s,t\}$,
\begin{align}
    p(a,b|x,y)
    =\sum_{\lambda,m}\pi(\lambda)
    p_{\mathrm e}(a,m|x,\lambda)
    p_{\mathrm d}(b|m,y,\lambda),
\end{align}
This is precisely the classical prepare-and-measure
strategy with a $d$-level message, as defined in
Eq.~(5). Thus, the quantum statistics generated
by the jointly measurable pair $(E_s,E_t)$
admit an exact classical simulation for both
measurement settings. Since $C_d(W^{st})$ is the
maximum of $W^{st}$ over all such classical
strategies, it follows that
\begin{align}
    W^{st}(E_s,E_t)\leq C_d(W^{st}).
\end{align}
\end{proof}
\begin{theorem}\label{Thm:two_setting}
Let $\cW$ be any linear figure of merit involving $n\geq 2$
measurement settings. For any parallel strategy involving two
receivers, any assemblage $\{\rho_{a|x}\}$ on $\mathbb{C}^d$,
and any broadcasting channel
$\Lambda\in\Chan(\mathbb{C}^d,\cH_{B_1}\otimes\cH_{B_2})$,
the figures of merit attained by the two receivers satisfy
\begin{equation}\label{Eq:two_setting_bound}
    \cW_1+\cW_2\leq 2\,\widetilde C_d(\cW).
\end{equation}
Consequently, Bob$_1$ and Bob$_2$ cannot simultaneously
attain values exceeding $\widetilde C_d(\cW)$. In particular,
for $n=2$, since $\widetilde C_d(\cW)=C_d(\cW)$, both receivers
cannot simultaneously exceed the classical bound $C_d(\cW)$.
\end{theorem}
\begin{proof}
For each receiver $k\in\{1,2\}$ and every pair of
measurement settings $s<t$, we have from Eq. \eqref{Eq:pair_merit} $\cW_k^{st}=V_s(E^k_s)+V_t(E^k_t)$. Since each $V_y$ appears in $n-1$ pairs, $\cW_k$ is as given in Eq.~\eqref{Eq:AverageMerit}. For a fixed pair, we exchange the $t$-th measurements of the two observers, which is allowed by the additivity~\eqref{Eq:additivity}. Thus,
\begin{align}
\cW^{st}_1+\cW^{st}_2
&=\big[V_s(E^1_s)+V_t(E^1_t)\big]+\big[V_s(E^2_s)+V_t(E^2_t)\big]\nonumber\\
&=\big[V_s(E^1_s)+V_t(E^2_t)\big]+\big[V_s(E^2_s)+V_t(E^1_t)\big]\nonumber\\
&=\cW^{st}(E^1_s,E^2_t)+\cW^{st}(E^2_s,E^1_t).\label{Eq:swap}
\end{align}
Both terms are values of $\cW^{st}$ for the \emph{same} assemblage, with measurements that come from \emph{different} observers. By Lemma~\ref{Lem:cross} the pairs $(E^1_s,E^2_t)$ and $(E^2_s,E^1_t)$ are jointly measurable, so by Lemma~\ref{Lem:JM_classical} each term is at most $C_d(\cW^{st})$. Therefore
\bea
\cW_1+\cW_2&=&\frac1{n-1}\sum_{s<t}\big(\cW^{st}_1+\cW^{st}_2\big)\nonumber\\
&&\le\frac{1}{n-1}\sum_{s<t}2\,C_d(\cW^{st})=2\,\widetilde C_d(\cW).\nonumber\\
\eea
If $\cW_1>\widetilde C_d(\cW)$ and $\cW_2>\widetilde C_d(\cW)$, the sum would exceed $2\widetilde C_d(\cW)$, a contradiction. The case $n=2$ follows directly from Lemma~\ref{Lem:Ctilde}.
\end{proof}

\begin{corollary}
For any parallel strategy involving $N\geq 2$
receivers, the figures of merit attained by any
two distinct receivers $i\neq j$ satisfy $\cW_i+\cW_j\leq 2\widetilde C_d(\cW)$.
Consequently, at most one receiver can exceed $\widetilde C_d(\cW)$.
\end{corollary}
\begin{proof}
For any pair of receivers $i\neq j$, define
the reduced broadcasting channel
$\Lambda_{ij}:=\operatorname{Tr}_{\overline{ij}}\circ\Lambda$,
where $\operatorname{Tr}_{\overline{ij}}$ denotes the
partial trace over all output systems except those
of receivers $i$ and $j$.
Since $\Lambda_{ij}$ preserves the marginal
statistics and hence the figures of merit
$\cW_i$ and $\cW_j$, applying Theorem~2 to
this reduced two-receiver strategy gives $\cW_i+\cW_j\leq 2\widetilde C_d(\cW)$.
\end{proof}

\begin{remark}
\label{Rem:why_two}
The proof of Theorem~\ref{Thm:two_setting} relies on the joint
measurability of effective measurements belonging to different observers in the parallel strategy. For $n=2$, this allows the combined figure of merit to be bounded directly by the classical bound $C_d(\cW)$. However, for $n\geq3$, the same argument cannot be applied to the complete figure of merit, since in Eq. \eqref{Eq:swap} each term will contain at least two measurements of the same observer and measurements belonging to the same Bob need not be jointly measurable. The argument therefore applies only to pairs of measurement settings, and in general the for $n\geq3$ the bound is given by  $\widetilde C_d(\cW)$. In contrast, in a sequential strategy, Bob$_1$'s measurement
generally disturbs the system subsequently received by
Bob$_2$. Although Bob$_1$'s measurement is jointly measurable
with the effective measurement of Bob$_2$ for a fixed choice
of $y_1$, this compatibility need not hold when Bob$_1$'s
setting is averaged over. Consequently, the joint
measurability condition of Lemma~\ref{Lem:cross} does not
generally hold for sequential strategies, and the argument
establishing Theorem~\ref{Thm:two_setting} is no longer
applicable.
\end{remark}

Remark \ref{Rem:why_two} suggests the possibility of sharing quantum advantage among multiple observers through a parallel strategy in communication tasks involving more than three decoding measurement settings for each observer. We explore this possibility in detail through specific communication tasks discussed in the following sections.
%==========================================================================
\section{Implication to two party communication task: random access codes}\label{Sec:PM_RAC}

To illustrate how the linear figure of merit introduced in Sec.~\ref{Sec:two_setting} arises in a specific communication task, namely, the RAC, we first set $a=1$, rendering Alice's outcome trivial, so that $\sigma_{1|x}=\rho_x\in\cS(\mathbb{C}^d)$ represent the prepared states. We drop the index $a$, so that the Eqs.~\eqref{c_stat} and \eqref{q_stat} become $p(b|x,y)=\sum_mp_e(m|x)p_d(b|m,y)$ and $p(b|x,y)=\tr[\rho_xM_y(b)]$ respectively. This is the standard prepare-and-measure scenario. With this modification the linear figure of merit becomes
\begin{align}
 &\cW[p]=\sum_{y\in[n]}V_y(F),\nonumber\\
\text{with \ } &V_y(F)=\sum_{x\in[X],b\in[B]}c(x,y,b)\tr\big[\rho_xF(b)\big] \nonumber,   
\end{align}
 where $p(b|x,y)=\tr\big[\rho_xF(b)\big]$, and the classical bound $C_d(\cW)$ is the maximum over all classical strategies with a $d$-level message and shared randomness. All results of Sec.~\ref{Sec:two_setting} apply with this specialization. In this scenario, we have the following observation.
\begin{observation}\label{Cor:PM}
In this scenario, for every linear $\cW$, every set of states $\{\rho_x\}\subset\cS(\mathbb{C}^d)$, every channel $\Lambda$ and all measurements of the receivers, $\cW_i+\cW_j\le2\widetilde C_d(\cW)$ for all $i\ne j$. For $n=2$ settings this reads $\cW_i+\cW_j\le2C_d(\cW)$, so at most one receiver can exceed the classical bound.
\end{observation}
With this observation in hand, we now explicitly investigate the possibility of sharing quantum advantage in RACs through a parallel strategy.
\subsection{Which random access codes cannot be shared in parallel?}\label{Sec:RAC_classification}
Consider an $(n,d)$-random access code (RAC), in which Alice receives a string $x=x_1\cdots x_n$ drawn uniformly at random from $[d]^n$. The receiver receives the input $y$ chosen uniformly from $[n]$ and must guess $x_y$ correctly. The average success probability for this task is then
\begin{equation}\label{Snd}
\mathcal S(n,d)=\frac1{nd^n}\sum_{x,y}p(b=x_y|x,y).
\end{equation}
This is a linear merit with $c(x,y,b)=\frac1{nd^n}\delta_{b,x_y}$, so
\begin{equation}
V_y(F)=\frac1{nd^n}\sum_x\tr\big[\rho_xF(x_y)\big],\qquad
\cW^{st}=\frac1n\big[\mathcal S(s)+\mathcal S(t)\big],
\end{equation}
where $\mathcal S(y):=d^{-n}\sum_xp(b=x_y|x,y)$, so that $\mathcal S(n,d)=\frac1n\sum_y\mathcal S(y)$. Up to the factor $\frac1n$, $\cW^{st}$ is the $(2,d)$-RAC on $(x_s,x_t)$ with the other dits ignored. We write $\mathcal S_i$ for the value of Bob$_i$ and $\mathcal S_C(n,d):=C_d(\cW)$.

\begin{lemma}\label{Lem:pair_RAC}
For all $n\geq2$, $d\geq2$, and any two distinct
measurement settings $s\ne t$, the classical bound
of the two-setting figure of merit satisfies
\begin{equation}
    C_d(\cW^{st})=\frac1n\Big(1+\frac1d\Big).
\end{equation}
\end{lemma}

\begin{proof}
Since the figure of merit is linear, its classical
maximum is attained by a deterministic strategy.
Let Alice encode the input string $x\in[d]^n$
into a message $m=\mu(x)\in[d]$, and let Bob's
guess for setting $y$ be $g_y(m)\in[d]$.
For the two settings $s$ and $t$, we define $
 \varsigma(x):=
    \delta_{g_s(\mu(x)),x_s}
    +\delta_{g_t(\mu(x)),x_t}
    \in\{0,1,2\}$.
Thus, $\varsigma(x)$ counts the number of correct
guesses for the input string $x$, and $
    \mathcal S(s)+\mathcal S(t)
    =d^{-n}\sum_x\varsigma(x)$.
Now, $\varsigma(x)=2$ only if both guesses are
correct, which requires
\begin{align}
    (x_s,x_t)\in
    T:=\{(g_s(m),g_t(m)):m\in[d]\}.
\end{align}
Since there are only $d$ possible messages,
$|T|\leq d$. Moreover, each fixed pair
$(x_s,x_t)$ corresponds to exactly $d^{n-2}$
input strings, as the remaining $n-2$ entries
can be chosen arbitrarily. Therefore, at most
$d\cdot d^{n-2}=d^{n-1}$ input strings can
satisfy $\varsigma(x)=2$.

Using $\varsigma(x)\leq
1+\delta_{\varsigma(x),2}$, we obtain,

\begin{align}
    \sum_x\varsigma(x)
    &\leq d^n+\sum_x\delta_{\varsigma(x),2} \nonumber\\
    &\leq d^n+d^{n-1}.
\end{align}
Consequently we have,
\begin{align}
    \mathcal S(s)+\mathcal S(t)
    \leq 1+\frac1d.
\end{align}
This bound is attainable by choosing
$\mu(x)=x_s$, $g_s=\mathrm{id}$, and
$g_t=\mathrm{const}$. Indeed, this strategy
gives $\mathcal S(s)=1$ and
$\mathcal S(t)=1/d$. Hence the maximum of
$\mathcal S(s)+\mathcal S(t)$ is exactly
$1+1/d$. Finally, since
$\cW^{st}=[\mathcal S(s)+\mathcal S(t)]/n$,
we conclude that
\begin{align}
    C_d(\cW^{st})
    =\frac1n\Big(1+\frac1d\Big).
\end{align}
\end{proof}

Lemma~\ref{Lem:pair_RAC} gives $\widetilde C_d(\cW)=\frac1{n-1}\binom n2\frac1n\big(1+\frac1d\big)=\frac12\big(1+\frac1d\big)$ for every $n\ge2$.

\begin{proposition}\label{Prop:RAC_class}
In every parallel strategy for the $(n,d)$-RAC, for all $n\ge2$ and $d\ge2$,
\begin{equation}
\mathcal S_1+\mathcal S_2\le1+\frac1d,
\end{equation}
and for $N$ receivers $\mathcal S_i+\mathcal S_j\le1+\frac1d$ for all $i\ne j$. Moreover, $\mathcal S_C(n,d)\le\frac12\big(1+\frac1d\big)$, with equality in the following cases:
\begin{enumerate}
\item[(i)] $n=2$ and any $d$ (Alice sends $x_1$);
\item[(ii)] $(n,d)=(3,2)$ (Alice sends the majority bit, giving $\mathcal S_C=\frac34$).
\end{enumerate}
In these cases at most one receiver can exceed $\mathcal S_C(n,d)$, so the quantum advantage cannot be shared in parallel.
\end{proposition}
\begin{proof}
The first statement is Theorem~\ref{Thm:two_setting}  with $\widetilde C_d(\cW)=\frac12(1+\frac1d)$. The bound $\mathcal S_C\le\widetilde C_d$ is Lemma~\ref{Lem:Ctilde}. For (i), sending $x_1$ gives $\mathcal S=\frac12(1+\frac1d)$. For (ii), the majority strategy gives $\frac34=\frac12(1+\frac12)$. In both cases $2\mathcal S_C=1+\frac1d$, so $\mathcal S_1>\mathcal S_C$ and $\mathcal S_2>\mathcal S_C$ would give $\mathcal S_1+\mathcal S_2>1+\frac1d$, which is impossible.
\end{proof}

Note that, (a) For $d=2$ the classical value is $\mathcal S_C(n,2)=\frac12+\binom{n-1}{\lfloor(n-1)/2\rfloor}2^{-n}$, which gives $\frac34,\frac34,\frac{11}{16},\frac{11}{16}$ for $n=2,3,4,5$. Hence for $n\ge4$ the window $\big(\mathcal S_C(n,2),\frac34\big)$ is not excluded by Proposition~\ref{Prop:RAC_class}.
(b) For $n\ge3$ the strategy ``send $x_1$'' gives $\frac1n\big(1+\frac{n-1}d\big)$, which is strictly below $\frac12\big(1+\frac1d\big)$ for $d\ge2$. For $(n,d)=(3,2)$ the majority strategy closes the gap. For $d\ge3$ we do not determine $\mathcal S_C(n,d)$ here. 

\begin{remark}
By Proposition~\ref{Prop:RAC_class}, the quantum advantage of the $(2,2)$-RAC cannot be shared in parallel by more than one receiver. In contrast, Ref.~\cite{sasmal2026} constructs a set of instruments by which an arbitrary number of receivers can each obtain an advantage in the $(2,2)$-RAC sequentially. This shows an arbitrarily large advantage of the sequential strategy over the parallel strategy.   
\end{remark}
Note that, for $n>2$, since the combined figure of merit is not necessarily bounded by the classical value $C_d$, the possibility of sharing quantum advantage remains open. We explore this possibility through a variant of RAC discussed in the following subsection.

\subsection{Sharing quantum advantage to multiple parties using universal cloning in $(3,2)$-PORAC}

An $(n,d)$-parity-oblivious random access code (PORAC) is a constrained variant of a standard RAC that retains the same winning condition while imposing an additional parity-obliviousness constraint on the information communicated between the observers. The additional operational constraint requires that no information about the parity of any subset containing two or more input bits can be transmitted to Bob. For a binary string $s\in\{0,1\}^n$ with Hamming weight $|s|\geq2$, define the parity $s\cdot x=\bigoplus_{j=1}^{n}s_jx_j$, where $\oplus$ denotes addition modulo two. The parity-obliviousness condition requires
\begin{align}
\frac{1}{2^{n-1}}
\sum_{x:s\cdot x=0}p(b\mid x,y)
=
\frac{1}{2^{n-1}}
\sum_{x:s\cdot x=1}p(b\mid x,y),
\end{align}
for every such $s$ and every measurement setting $y$ and outcome $b$. In quantum theory, where Alice encodes $x$ into a density operator $\rho_x$, this condition is satisfied if
\begin{align}
\frac{1}{2^{n-1}}
\sum_{x:s\cdot x=0}\rho_x
=
\frac{1}{2^{n-1}}
\sum_{x:s\cdot x=1}\rho_x.
\end{align}
The maximum classical average success probability under the parity-obliviousness constraint is
\begin{align}
P_{\mathrm C}^{\mathrm{PORAC}}
=\frac12\left(1+\frac1n\right).
\end{align}

Now, Consider $(3,2)$-PORAC \cite{spekkens2009} whose classical bound is $P_{\mathrm C}^{\mathrm{PORAC}}\leq 2/3$. In the quantum version of the task Alice receives a uniformly distributed three-bit string
$x=(x_1,x_2,x_3)\in\{0,1\}^3$ and encodes it into a qubit state
\begin{equation}
    \rho_x=\frac{1}{2}\left(I+\vec{r}_x\cdot\vec{\sigma}\right),
\end{equation}
with $\vec{r}_x=\frac{1}{\sqrt{3}}\left((-1)^{x_1},(-1)^{x_2},(-1)^{x_3}\right)$ being the Bloch vector. Alice applies an asymmetric universal $1 \rightarrow 2$ quantum cloning channel $\Lambda$ to the prepared state $\rho_x$ and distributes the two output systems to Bob$_1$ and Bob$_2$, respectively. The resulting bipartite state is $\rho_x^{B_1B_2}=\Lambda(\rho_x)$. Then the two systems to be sent to Bob$_1$ and Bob$_2$ are obtained from the $\rho_x^{B_1B_2}$ as,
\begin{align}
\label{eq:Density_prep3-bit}
    \rho_x^{B_1}
    &=\Tr_{B_2}[\Lambda(\rho_x)], \ \text{and} \ \ \rho_x^{B_2} =\Tr_{B_1}[\Lambda(\rho_x)].
\end{align}

For an asymmetric universal cloning transformation, the two reduced states take the form \cite{Bruss97}
\begin{align}
\label{eq:Channelreduced}
    \rho_x^{B_1}
    =\eta_1\rho_x+(1-\eta_1)\frac{I}{2} \ , \ \ \
    \rho_x^{B_2}
    =\eta_2\rho_x+(1-\eta_2)\frac{I}{2} \ ,
\end{align}
where the shrinking factors are parametrized as
\begin{align}
    \eta_1=1-\beta^2,\qquad
    \eta_2=1-\alpha^2,
\end{align}
subject to
\begin{align}
    \alpha^2+\beta^2+\alpha\beta=1,
    \qquad \alpha,\beta\geq 0.
\end{align}
Further, substituting Eq. \eqref{eq:Density_prep3-bit} into Eq. \eqref{eq:Channelreduced} we get,
\begin{align}
    \rho_x^{B_1}
    =\frac{1}{2}
    \left(I+\eta_1\vec{r}_x\cdot\vec{\sigma}\right),\ \
    \rho_x^{B_2}
    =\frac{1}{2}
    \left(I+\eta_2\vec{r}_x\cdot\vec{\sigma}\right).
\end{align}
Now, upon receiving an independently and uniformly chosen input
$y_i\in\{1,2,3\}$, Bob$_i$ performs the two-outcome projective
measurement represented as,
\begin{align}
    M_{b_i|y_i}^{(i)}
    =\frac{1}{2}
    \left(I+(-1)^{b_i}\sigma_{y_i}\right),
    \qquad b_i\in\{0,1\},
\end{align}
where $\sigma_1=\sigma_x$, $\sigma_2=\sigma_y$, and
$\sigma_3=\sigma_z$. The objective of Bob$_i$ is to output
$b_i=x_{y_i}$ in each run of experiment according to his input $y$. Therefore, the average success probability is
\begin{align}
\label{eq:3bitProb}
    P_i
    =\frac{1}{24}
    \sum_{x\in\{0,1\}^3}
    \sum_{y_i=1}^{3}
    \operatorname{Tr}
    \left[
    \rho_x^{B_i}M_{x_{y_i}|y_i}^{(i)}
    \right].
\end{align}

Substituting the reduced cloned states and measurements in Eq. \eqref{eq:3bitProb} we get,

\begin{align}
    P_i
    &=\frac{1}{24}\sum_{x,y_i}
    \operatorname{Tr}\left[
    \frac{1}{2}
    (I+\eta_i\vec{r}_x\cdot\vec{\sigma})
    \frac{1}{2}
    (I+(-1)^{x_{y_i}}\sigma_{y_i})
    \right]\nonumber\\
    &=\frac{1}{96}\sum_{x,y_i}
    \operatorname{Tr}\Big[
    I+(-1)^{x_{y_i}}\sigma_{y_i}
    +\eta_i\vec{r}_x\cdot\vec{\sigma}
    \nonumber\\
    &\hspace{35mm}
    +\eta_i(-1)^{x_{y_i}}
    (\vec{r}_x\cdot\vec{\sigma})\sigma_{y_i}
    \Big].
\end{align}

This on simplification gives,
\begin{align}
\label{eq:3bitProb2}
    P_i
    &=\frac{1}{96}\sum_{x,y_i}
    \left[
    2+2\eta_i(-1)^{x_{y_i}}r_{x,y_i}
    \right]\nonumber\\
    &=\frac{1}{48}\sum_{x,y_i}
    \left[
    1+\eta_i(-1)^{x_{y_i}}r_{x,y_i}
    \right].
\end{align}
Where we have used, Using the Pauli matrix identities $\Tr[I]=2$, $ \Tr[\sigma_j]=0 $, and, $\Tr[\sigma_j\sigma_k]=2\delta_{jk}$.

It is known that Alice's encoding is $ r_{x,y_i}=\frac{(-1)^{x_{y_i}}}{\sqrt{3}}$, and, hence,
\begin{align}
    (-1)^{x_{y_i}}r_{x,y_i}
    =\frac{(-1)^{2x_{y_i}}}{\sqrt{3}}
    =\frac{1}{\sqrt{3}}.
\end{align}

Therefore, from Eq. \eqref{eq:3bitProb2} the optimal success probability of the PORAC task performed by Alice-Bob$_i$ becomes,
\begin{align}
    P_i
    =\frac{1}{48}\sum_{x,y_i}
    \left(1+\frac{\eta_i}{\sqrt{3}}\right)=\frac{1}{2}
    \left(1+\frac{\eta_i}{\sqrt{3}}\right).
\end{align}

Finally, substituting the shrinking factors $\eta_1$ and $\eta_2$ of the asymmetric universal cloning transformation, we obtain the success probabilities corresponding to Alice-Bob$_1$ and Alice-Bob$_2$ as,
\begin{align}
\label{eq:3bitProbFinal}
    P_1=\frac{1}{2}
    \left(1+\frac{1-\beta^2}{\sqrt{3}}\right),\ \ \ P_2=\frac{1}{2}\left(1+\frac{1-\alpha^2}{\sqrt{3}}\right),
\end{align}

To obtain quantum advantage in the three-bit PORAC simultaneously for both the Bobs, both must exceed the classical bound $P_{\mathrm C}=2/3$. Thus, the conditions for simultaneous advantage in parallel sharing by using cloning protocol are
\begin{align}
    P_1>\frac{2}{3}
    &\iff 1-\beta^2>\frac{1}{\sqrt{3}},\\
    P_2>\frac{2}{3}
    &\iff 1-\alpha^2>\frac{1}{\sqrt{3}}.
\end{align}
Equivalently we may write the conditions as,
\begin{align}
    \alpha^2<1-\frac{1}{\sqrt{3}},
    \qquad
    \beta^2<1-\frac{1}{\sqrt{3}}.
\end{align}

These inequalities admit simultaneous solutions consistent with the cloning constraint. In particular, the symmetric universal cloner is recovered as the special case $\alpha=\beta=1/\sqrt{3}$, giving
\begin{align}
    P_1=P_2
    =\frac{1}{2}
    \left(1+\frac{2}{3\sqrt{3}}\right)
    \approx 0.69245>\frac{2}{3}.
\end{align}

\begin{figure}
     \includegraphics[height=70mm, width=95mm,scale=1.5]{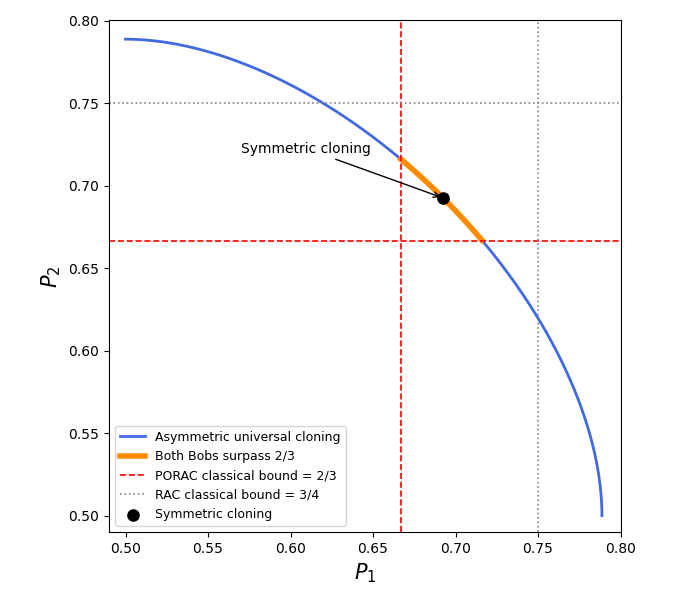}
    \caption{Optimal Tradeoff between the success probabilities achieved by of Bob$_1$ and Bob$_2$ in $3$-bit PORAC task.}
    \label{Q-tradeoff}
\end{figure}

The general trade-off between the success probabilities of the two observers is illustrated in self explanatory Fig.~\ref{Q-tradeoff}. Therefore, asymmetric universal quantum cloning allows two spatially separated Bobs to simultaneously obtain quantum advantage in a three-bit PORAC, without requiring sequential measurements or classical communication between the receivers.

\section{Concluding Remarks}\label{sec:concl}
In this work, we have investigated the possibilities and limitations of sharing quantum advantage among multiple observers through parallel and sequential strategies in prepare-and-measure communication tasks. While sequential sharing relies on successive interactions with a quantum system, parallel sharing employs quantum broadcasting to distribute information among multiple receivers. We have examined the relationship between these strategies in terms of the statistics they can reproduce and their ability to distribute quantum advantage in prepare-and-measure communication tasks.

We have established that any statistics achievable through a parallel strategy involving two observers can also be reproduced sequentially, whereas this correspondence does not generally hold for more than two observers. We have further derived a general restriction on sharing quantum advantage through parallel strategies for communication tasks characterized by linear figures of merit. In particular, we have shown that simultaneous quantum advantage is impossible for tasks involving two decoding measurements. Furthermore, it is demonstrated that, although parallel sharing of quantum advantage is impossible in certain random access codes, it becomes possible in a three-bit parity-oblivious random access code, where two observers can simultaneously outperform the classical limit using imperfect quantum clones. These results highlight the fundamental differences between parallel and sequential information sharing and identify the role of communication-task constraints in determining the possibility of distributing quantum advantage.

Several open questions emerge from our work. First, it would be interesting to characterize broadcasting channels in terms of their ability to distribute nonclassicality among multiple receivers. Such a characterization could help identify broader classes of communication tasks \cite{GISIN199713,clone06,clone01,cloneCodes} in which parallel sharing of quantum advantage is possible. Another promising direction is to investigate quantum interactions from the perspective of sustainable information sharing, particularly the trade-off between information extraction and the preservation of quantum resources for subsequent observer \cite{ghai2026information,ghaiInst26}. It would also be worthwhile to extend our analysis beyond dimensional restrictions by considering other resource constraints, such as bounds on information and energy. Finally, exploring the potential applications of these sharing strategies in information-theoretic and computational tasks may offer further insights into the possibilities and limitations of distributed quantum information processing.
\section*{Acknowledgement}
AM acknowledges STARS (Grant No. STARS/STARS-2/2023-0809), Government of India, for support, and, SM acknowledges ANRF National Post-Doctoral Fellowship (NPDF) under File No. PDF/2025/000262.
\section*{AI Usage Statement}
The central idea of this work was conceived by the authors, who also performed all the principal calculations and derivations. The large language model Claude (Anthropic) was used to assist in checking the proofs, particularly in formalizing Theorem~\ref{Thm:no_dim_d}, and improving the language and phrasing of the manuscript. 
\bibliography{reference}
\end{document}